\documentclass[12pt,a4paper]{article}%
\usepackage{jheppub}
\pdfoutput=1
\usepackage{amsmath,amssymb,amsfonts,wasysym}
\usepackage[bf,footnotesize]{caption2}
\usepackage[]{hyperref}%
\hypersetup{
colorlinks=true,
linkcolor=red,
anchorcolor=black,
citecolor=blue,
filecolor=cyan,
menucolor=red,
runcolor=filecolor,
urlcolor=magenta,
bookmarks=true,
bookmarksnumbered=true
}
\def\l{\label}
\def\La{\mathcal{L}}

\def\({\left(}
\def\){\right)}
\def\f{\frac}
\def\be{\begin{equation}}
\def\ee{\end{equation}}
\def\s{\sigma}

\def\ovl{\overline}

\title{Limits on leptophobic $W'$ after 1 fb$^{-1}$ of LHC data: a lesson on parton level simulations}
\author{Riccardo Torre}
\affiliation{Dipartimento di Fisica, Universit\`a di Pisa, Largo Fibonacci 3, I-56127 Pisa, Italy}
\affiliation{Institut f\"ur Theoretische Physik, Universit\"at Z\"urich, Winterthurerstrasse 190, CH-8057 Z\"urich, Switzerland}
\emailAdd{Riccardo.Torre@cern.ch}

\proceeding{}

\proceeding{\vspace{-1.65cm}\flushright{\small IFUP-TH/2011-18\\}\vspace{5mm}
\flushleft{\footnotesize Based on a talk given at the workshop\\\vspace{2mm}
23rd Rencontres de Blois\\
Ch\^ateau Royal de Blois, France, 01 June 2011}}

\abstract{An iso-singlet $W'$ carrying hypercharge one is considered within an effective Lagrangian approach. The predictions for the exclusion in the $\(M_{W'},g_{q}/g\)$ parameter space presented in Ref.~\cite{Grojean:2011vu} for the search in the di-jet invariant mass spectrum are shown to be in reasonable agreement with the LHC results based on 1 fb$^{-1}$ of integrated luminosity.}

\begin{document}
%
\maketitle


\section{Introduction}\l{sec1}
The possibility for the LHC to observe new physics at the TeV scale deserves the highest consideration. In particular, many Beyond the Standard Model (BSM) theories predict new particles around and above the Fermi scale $v\approx 246$ GeV. Here we consider the possibility that a new charged vector, generically referred to as a $W'$, exists with a mass around $1$ TeV. We employ an effective Lagrangian approach to describe the interactions of the new vector with the Standard Model (SM) particles only assuming its quantum numbers under the SM gauge group $G_{\text{SM}}=SU\(3\)_{C}\times SU\(2\)_{L}\times U\(1\)_{Y}$. In particular, in order to avoid constraints coming from an associated neutral $Z'$, we consider the $\(1,1\)_{1}$ representation of $G_{\text{SM}}$, which contains a charged vector with hypercharge $Y=1$ and no neutral states. The new charged vector can couple directly to quarks, in which case it is likely to be almost elementary\footnote{The case in which the third generation quarks are composite objects can be an exception.}, but can also have suppressed coupling to fermions and enhanced couplings to pairs of gauge bosons. In particular the $WZ$ and $W\gamma$ decay channels become the most relevant in the case of a composite object arising from an unknown strong dynamics, possibly related to the mechanism which realizes the ElectroWeak Symmetry Breaking (EWSB). Due to its iso-singlet nature the couplings to leptons of the new charged vector are usually strongly suppressed, making it generally less constrained than its iso-triplet counterpart.\\
This paper is organized as follows. In Section \ref{sec2} we briefly remind the general effective Lagrangian which describes the new charged vector and in Section \ref{sec3} we compare the predictions of Ref.~\cite{Grojean:2011vu} with the new bounds coming from the ATLAS and CMS searches for resonances in the di-jet invariant mass spectrum with $1$ fb$^{-1}$ of integrated luminosity. We conclude in Section \ref{sec4}.

\section{Effective Lagrangian approach}\l{sec2}
A general effective Lagrangian describing a new charged vector $V^{\pm}_{\mu}$ transforming in the $\(1,1\)_{1}$ representation of $G_{\text{SM}}$ is given by (see Ref.~\cite{Grojean:2011vu} for details)
\be\l{eq1}
\La=\La_{\text{SM}}+\La_{V}+\La_{V-\text{SM}}\,,
\ee
where $\La_{\text{SM}}$ is the SM Lagrangian, and 
\be\l{eq2}
\La_{V}=D_{\mu}V_{\nu}^{-}D^{\nu}V^{+\mu}-D_{\mu}V_{\nu}^{-}D^{\mu}V^{+\nu}+\tilde{M}^{2}V^{+\mu}V_{\mu}^{-}+\f{g_{4}^{2}}{2}\left|H\right|^{2}V^{+\mu}V_{\mu}^{-}-ig'c_{B}B_{\mu\nu}V_{\mu}^{+}V_{\nu}^{-}\,,
\ee
\vspace{-2mm}
\be\l{eq3}
\La_{V-\text{SM}}=V^{+\mu}\(ig_{H}H^{\dag}\(D_{\mu}\tilde{H}\)+\f{g_{q}}{\sqrt{2}}\(V_{R}\)_{ij}\ovl{u^{i}_{R}}\gamma_{\mu}d^{j}_{R}\)+\text{h.c.}\,,
\ee
where we have defined $\tilde{H}\equiv i\s^{2}H^{*}$.
This effective Lagrangian, which is assumed to be written in the mass eigenstate basis for fermions, contains five new parameters and the new mixing matrix $V_{R}$ in the right-handed (RH) sector. This matrix is arbitrary in the effective approach but the weakest bound coming from flavor physics ($K$ and $B$ meson mixings) is obtained for $\left|V_{R}\right|=\mathbf{I}_{3}$ \cite{Langacker:1989p2578}. We assume this particular form of $V_{R}$ in our phenomenological study. The other five parameters are given by the mass parameter $\tilde{M}$, the coupling $g_{4}$ of two $V_{\mu}$'s to two Higgses, the coupling $g_{H}$ of one vector to two Higgses, the coupling $g_{q}$ to the RH quark current and the coupling $c_{B}$ of two vectors to $B_{\mu\nu}$ which generates a $W'W\gamma$ coupling and modifies the $W'WZ$ coupling coming from the kinetic term in Eq.~\eqref{eq2} and from the Yang--Mills Lagrangian of the SM.\\
Upon EWSB the coupling $g_{H}$ generates a mass mixing between $V_{\mu}^{\pm}$ and the charged EW gauge boson $\hat{W}_{\mu}^{\pm}$ which can be diagonalized to find the mass eigenstates
\be\l{eq4}
\begin{pmatrix} W^{+}_{\mu} \\
W^{\prime\,+}_{\mu} \end{pmatrix} = \begin{pmatrix}
\cos\hat\theta\,\,\, & \sin\hat\theta \\
-\sin\hat\theta\,\,\, & \cos\hat\theta 
\end{pmatrix} \begin{pmatrix}
\hat{W}_{\mu}^{+} \\
V_{\mu}^{+}
\end{pmatrix}\,.
\ee
The mixing angle $\hat{\theta}$ is given by the expression
\be\l{eq5}
\tan(2\hat{\theta})=\frac{2\Delta^{2}}{m^{2}_{\hat{W}}-M^{2}}\,,
\ee
where 
\be\l{eq6}
m_{\hat{W}}^{2}=\frac{g^{2}v^{2}}{4},\quad \Delta^{2}=\frac{g_{H}gv^{2}}{2\sqrt{2}},\quad M^{2}=\tilde{M}^{2}+\frac{g_{4}^{2}v^{2}}{4}\,.
\ee
Further details on the model, including the explicit form of the trilinear couplings with the EW gauge bosons, can be found in Ref.~\cite{Grojean:2011vu}.

\section{Bounds on the coupling to quarks from di-jet searches}
\begin{figure}[h]
\begin{center}
\includegraphics[scale=0.33]{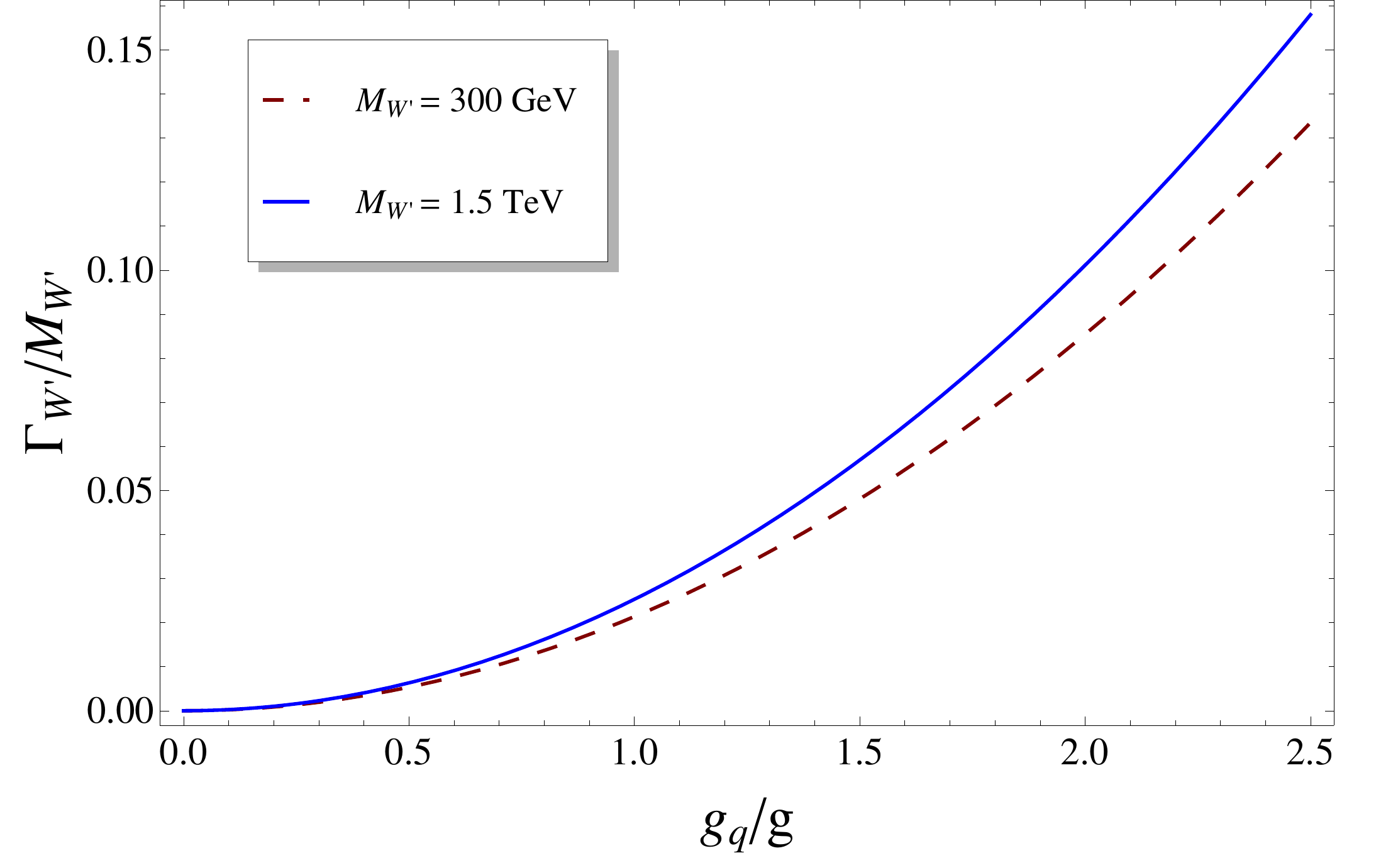}
\hspace{-1mm}\includegraphics[scale=0.33]{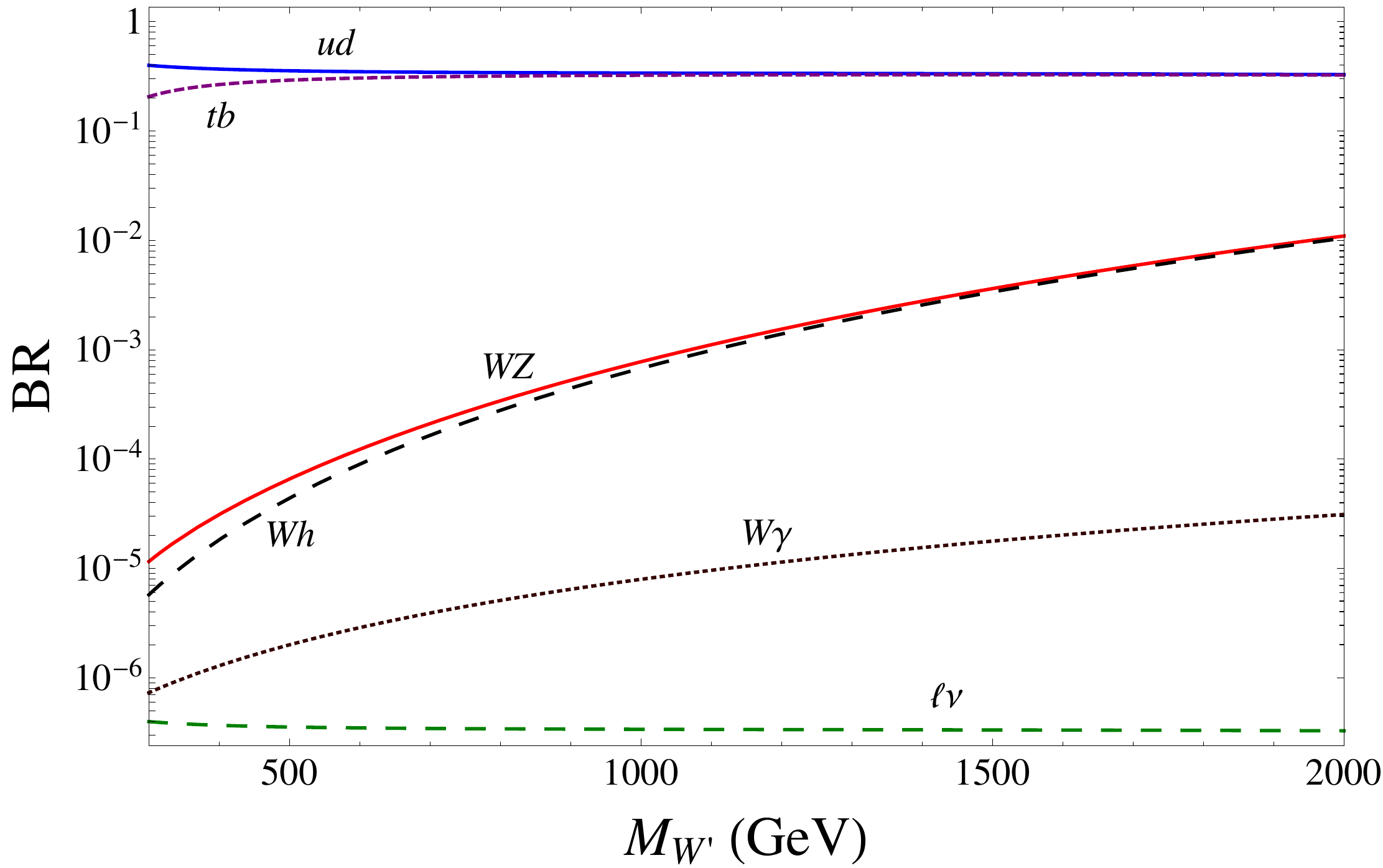}
\caption{\textsl{Left panel.} $W'$ width over mass ratio as a function of $g_{q}/g$  for negligible mixing, $\hat\theta \approx 0$, for $M_{W'}=300\,\mathrm{GeV}$ (dashed, red) and 1.5 TeV (blue). \textsl{Right panel.} Branching ratios of the $W'$ as a function of its mass, for the following choice of the remaining parameters: $g_{q}=g$, $\hat\theta=10^{-3}$, $c_{B}=-3$, $g_{4}=g$. From top to bottom: $ud$, $tb$, $WZ$, $Wh$, $W\gamma$, $\ell\nu$ (the latter includes all the three lepton families).}
\label{fig:totalwidth}
\end{center}
\end{figure}

\l{sec3}
Bounds of different nature on the parameters appearing in the Lagrangian \eqref{eq1} have been extensively discussed in Ref.~\cite{Grojean:2011vu}. Here we are interested in the bounds coming from direct searches in the di-jet invariant mass spectrum and we can therefore assume a negligible mixing angle $\hat{\theta}\approx 0$. The dependence of the ratio $\Gamma_{W'}/M_{W'}$ on the coupling $g_{q}$ is plotted in the left panel of Fig.~\ref{fig:totalwidth}, while the branching ratios as functions of $M_{W'}$ are shown in the right panel of the same figure, for representative values of the parameters.

\begin{figure}[h!]
\begin{center}
\includegraphics[scale=0.46]{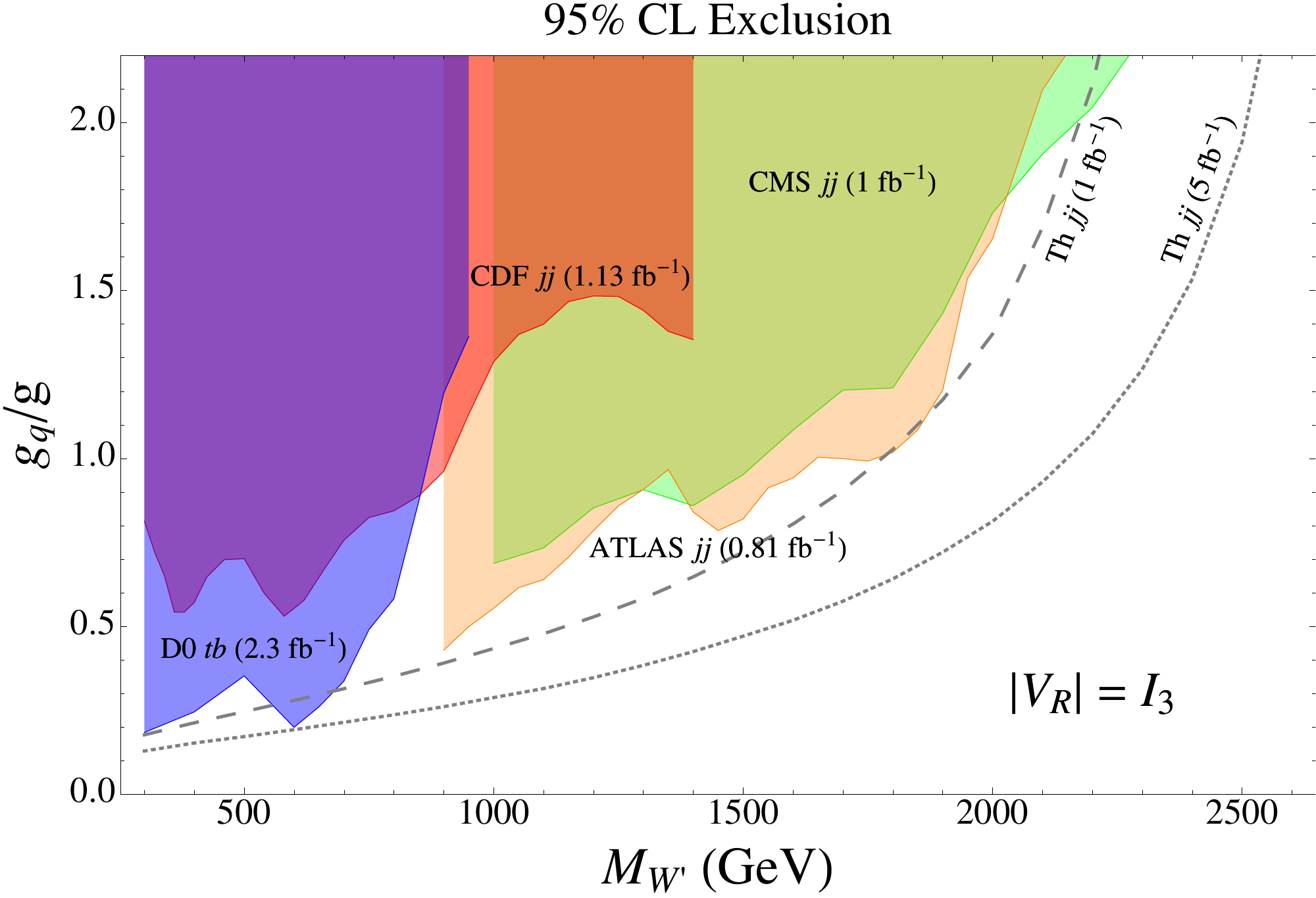}
\caption{Regions of the $\(M_{W'},g_{q}/g\)$ plane excluded at $95\%$ CL by the Tevatron and LHC searches in the di-jet final state (red region $0.3-1.4$ TeV for CDF di-jet search, orange region $0.9-2.1$ TeV for ATLAS di-jet search, green region $1-2.5$ TeV for CMS di-jet search) and the D0 search in the $tb$ final state (blue region $0.3-0.95$ TeV). Also shown in grey are the $95\%$ CL exclusion contours computed with a parton level simulation for an integrated luminosity of $1$ fb$^{-1}$ (dashed) and $5$ fb$^{-1}$ (dotted). For further details see Ref.~\cite{Grojean:2011vu}.}
\label{fig:Bounds}
\end{center}
\end{figure}

The most recent di-jet search at the Tevatron, based on 1.13 fb$^{-1}$ of data, has been performed by the CDF Collaboration \cite{CDFCollaboration:2009p2710} while the most recent search into the $tb$ final state at the Tevatron has been performed by the D0 Collaboration and is based on 2.3 fb${}^{-1}$ of data \cite{Abazov:2011xs}. The bounds coming from these analyses on the $(M_{W'},g_{q}/g)$ parameter space have been studied in Ref.~\cite{Grojean:2011vu} and are summarized in Figure \ref{fig:Bounds} (notice that the D0 bound has been updated with respect to Ref.~\cite{Grojean:2011vu}).

The ATLAS and CMS Collaborations have performed searches for resonances in the di-jet invariant mass spectrum with $1$ fb$^{-1}$ of integrated luminosity, respectively in Refs.~\cite{ATLASCollaboration:2011ww} and \cite{Chatrchyan:2011ns}. The bounds coming from these analyses are also shown in Figure \ref{fig:Bounds}. The analyses of the ATLAS and CMS Collaborations focus on high mass resonances with minimum masses of $0.9$ and $1$ TeV respectively. Moreover, the two limits have been set using different techniques. In particular, the CMS analysis takes into account the different quantity of QCD radiation generated by the $qq$, $qg$ and $gg$ final states, while the ATLAS analysis sets a model independent bound on the cross section times the acceptance as a function of the mass and the width of the new state assuming the signal shape to be gaussian. The different kinematic requirements of the two experiments are summarized in Table \ref{table1}. To take into account the readout problems of the ATLAS calorimeter in the region $-0.1<\eta_{j}<1.5$ and $-0.9<\phi_{j}<-0.5$ we have reduced the kinematic acceptance for the ATLAS kinematic requirements by a factor $0.9$ \cite{ATLASCollaboration:2011ww}. Further details on the two analyses can be found in Refs. \cite{ATLASCollaboration:2011ww} and \cite{Chatrchyan:2011ns}. All the bounds in Figure \ref{fig:Bounds} have been set comparing the experimental limits with a parton level simulation performed using the CalcHEP matrix element calculator \cite{Belyaev:Df-Kj9yc}. 
\begin{table}[t]
\begin{center}
\begin{tabular}{c|c|c}
{\bf Variable} 		& {\bf ATLAS} 	& {\bf CMS}\\ \hline
$M_{jj}$			&  --- 			&  $> 838$ GeV\\
$p_{T_{j}}$		&  $> 180$ GeV 	&  $> 717$ GeV\\
$|\eta_{j}|$		& $< 2.8$			& $< 2.5$\\
$|\Delta \eta_{jj}|$	& $< 1.2$			& $< 1.3$\\
\hline
\end{tabular}
\end{center}
\caption{\small\em Kinematic requirements used by the ATLAS and CMS analyses. The ATLAS analysis also excluded the region $-0.1<\eta_{j}<1.5$ and $-0.9<\phi_{j}<-0.5$, affected by a temporary readout problem \cite{ATLASCollaboration:2011ww}.}\label{table1}
\end{table}

From Figure \ref{fig:Bounds} we see that in spite of the different analysis techniques, the exclusion limits of the ATLAS and CMS Collaborations are compatible with each other. Moreover, both these limits are in reasonable agreement with the parton level prediction obtained in Ref.~\cite{Grojean:2011vu} and represented by the dashed grey line in Figure \ref{fig:Bounds}. This shows in particular that a reasonable prediction for the exclusion/discovery of a new physics model can be obtained with a simple parton level simulation provided that the signal and background distributions are integrated in order to weaken the dependence on detector effects and systematic uncertainties\footnote{For further details on the parton level simulation see Ref.~\cite{Grojean:2011vu}}.

\section{Conclusion}\l{sec4}
The effective model describing a new charged vector introduced in Ref.~\cite{Grojean:2011vu} was reconsidered and new bounds on the parameter space $\(M_{W'},g_{q}/g\)$ have been set using the recent analyses of the ATLAS and CMS Collaborations with $1$ fb$^{-1}$ respectively. The new bounds have been shown to be in good agreement with the predictions of Ref.~\cite{Grojean:2011vu} obtained with a simple parton level simulation.

\section*{Acknowledgments}
This project was done in collaboration with C.~Grojean and E.~Salvioni.\\
This work has been partially supported by the European Commission under the contract ERC Advanced Grant 226371 MassTeV, the contract PITN-GA-2009-237920 UNILHC and the contract PITN-GA-2010-264564 LHCPhenoNet and by the Swiss National Science Foundation under the Grant n. 200021-125237.

\def\bstname{fdp}

%
 \bibliographystyle{JHEP}
\bibliography{paper}{}
\end{document}